\documentclass[fleqn,12pt,twoside]{article}
\usepackage{espcrc1}
\usepackage{graphicx} 
\usepackage{epsfig}
\usepackage{subfigure}

\newcommand{\AmS}{{\protect\the\textfont2
  A\kern-.1667em\lower.5ex\hbox{M}\kern-.125emS}}

\title{Optical Calibration For Jefferson Lab HKS Spectrometer}
\author{L. Yuan and L. Tang\address[MCSD]{Department of Physics, \\ 
       Hampton University, Hampton, VA 23668}}

\begin{document}

\maketitle

\begin{abstract}
In order to accept very forward angle scattering particles, Jefferson
Lab HKS experiment uses an on-target zero degree dipole magnet.  The
usual spectrometer optics calibration procedure has to be modified due
to this on-target field.  This paper describes a new method to
calibrate HKS spectrometer system. The simulation of the calibration
procedure shows the required resolution can be achieved from initially
inaccurate optical description.
\end{abstract}
 
\section {Introduction}
 Jefferson Lab HKS experiment aims at obtaining high resolution
 hypernuclear spectroscopy (e.g., $\sim 400$ KeV FWHM for
 $^{12}_\Lambda$B) by $(e,e'K)$ reaction (ref.\cite{nuenpa}). It
 utilies CEBAF primary electron beam. The electroproduction provides a
 advantage over hadronic production mechanism because the momentum and
 position spread of the primary electron beam is significantly less
 than secondary hadron beams. The kaons are detected by High
 Resolution Kaon Spectrometer (HKS) with a momentum resolution of
 $2\times10^{-4}$. In order to maximize
 hypernuclear yield, the target is placed just before a zero degree
 dipole magnet (Splitter). The field of the splitter is used to
 seperate scattered e' from K$^+$ and allow the spectrometer to accept
 both e' and K$^+$ from very forward angles. 

To calculate hypernuclear excitation energies, we need to determine
scattered kaon and electron momenta and angles. The kaon
momentum is the dominant factor for the final missing mass
resolution. In order to obtain the proposed missing mass resolution,
it is important to optimize the spectrometer optics to get the
required momentum and angle reconstruction resolution. This
calibration procedure is complicated for HKS spectrometer Because of the
existence of splitter magnet right after the target.

\begin{table}[!hbt]
\caption{required momentum and angle reconstruction resolution}
\begin{center}
\begin{tabular}{ll} \hline
K+ momentum   & 110 KeV/c (rms):1$\times10^{-4}$ \\ 
E' momentum      & 120 KeV/c(rms):2$\times10^{-4}$  \\
K+ angle    & 2.9 mrad(rms)\\\hline
\end{tabular}
\end{center}
\end{table}

The spectrometer optics can be conviniently expressed in a matrix
formalism (ref.\cite{offermannnim}), which is a set of polynimial
coefficients related target angles and particle momentum to focal
plane measurements.  Conventionally, the method for calibrate matrix
for angle reconstruction is by using a special designed collimator
called sieve slit with a array of holes drilled onto metal template
(ref.\cite{offermannnim}). The sieve slit is mounted in front of the
spectrometer entrance so all the particles passing through the
holes will have defined incident angles.
 
\begin{figure}
\begin{center}
\includegraphics[width=9cm,bbllx=30, bblly=100,
  bburx=600,bbury=420,viewport=30 100 600 350,clip]{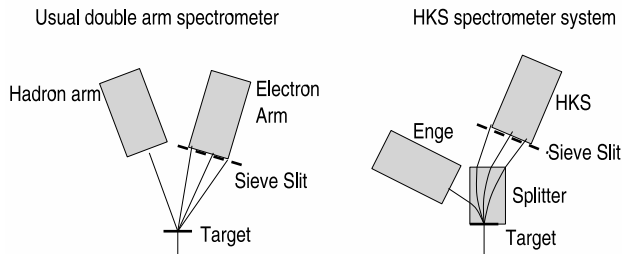}
\end{center}
\caption{Schematic view of magnetic optical properties of normal
  double arm spectrometer system (left) and HKS spectrometer
  (right) with an on-target dipole magnet (Splitter).}
\label{fig:hksoptics}
\end{figure}

But in the setup of HKS experiment, because target is inside the field
of Splitter, the sieve slit plate can only be mounted on the Enge and
HKS dipole entrance after the Splitter (fig.\ref{fig:hksoptics}). The
particle trajectories are already bended by the splitter field before
they pass through the sieve slit holes. Thus a hole of sieve slit no
longer select a single, fixed angle.  The usual way of angle
reconstruction calibration has to be modified.

For momentum calibration, because Enge, HKS two arms and the beam dump line
are all coupled through splitter field, thus suitable kinematics
setting for typical single arm momentum calibration (e.g., by
production of carbon excited states) is difficult to implement. 

In this paper a high precision optical calibration method for HKS
spectrometer is described. The procedure can be applied to optical
calibration of other magnet spectrometer system with on-target
magnetic field.  The optical calibration procedure will use the data
from Enge and HKS sieve slit runs for angle reconstruction calibration
and CH$_2$ target runs for momentum calibration. Before the
experiment, the procedure has been tested by simulation with initial
optics. The simulation is carried out by using measured field map of
the spectrometer magnets and tracking particles through the field by
program GEANT (ref. \cite{geantman}). The calibration procedure for
angle, momentum, kinematics reconstruction and beam raster correction
will be described from Sec.1 to Sec.5 seperately. The iteration of
these procedure is described in Sec.6. The simulation and result of
the calibration procedure from is presented in sec.7, followed by a
summary in section 8.

\section {Sieve Slit Calibration}

In HKS experiment, although the Splitter field smears the one to one
correspondence between target angles and sieve slit hole position. The
trajectory of a particle which passing through a given sieve slit hole
and scattered from a point target is a functional of the paticle
momentum and splitter optics. Thus with known splitter optics, the
scattering angle can be expressed as a function of particle momentum
and sieve slit hole position. The optics of Splitter can be obtained
by measuring the field distribution of the magnet and tracking
particles through the field by programs such as GEANT (ref.\cite{geantman}). From
the point of view of total integral of $Bdl$, the Splitter
contribution is only $\sim 8$\%, because the Splitter is just used to
seperate particles. The contribution of field measurement error is
negligible for target angle reconstruction.

 Based on this observation, the sieve slit(SS) calibration
procedure is:

\begin {enumerate}
\item Fit function $F_{s2t}$: target angles as a function of SS hole
positions and particle momentum based on Splitter optics.

\item Separate events of the SS calibration run data hole by hole,
thus determine corresponding SS hole center
position for each event.

\item Using function $F_{s2t}$, calculate corrected target angles
($xptar$, $yptar$) for each event from SS center position and
momentum.

\item Fit transfer matrix from focal plane to target: ($xfp$, $xpfp$, $yfp$,
$ypfp$)$\longrightarrow$($xptar$, $yptar$).
\end {enumerate}

A simulation of the SS calibration procedure with initial inaccurate
Splitter, Enge optics reaches final resolution for e' (Enge) target
angles ($\sigma$): $xptar$: 1.8 mr, $yptar$: 1.1 mr, for kaon (HKS)
angles: $xptar$: 0.4 mr, $yptar$: 2.4 mr.

\section{Momentum Calibration}

The method of HKS momentum calibration is making use of the known
masses of $\Lambda$,$\Sigma^0$ hyperons and the narrow width of
$^{12}_\Lambda$B hypernuclear ground state (ref.\cite{PDB}). The hyperons is produced
from hydrogens in the CH$_2$ foil (5 mm in thickness) The momentum
reconstruction matrices for Enge and HKS arms are optimized
simultaneously by minimizing the Chisquare defined as the sum of
squared mass differences between the calculated mass and the PDB
values. A Nonlinear Least Square method is used to optimize the
matrices for electron arm and kaon arm simutaneously.

Although there is emulsion measurement of $^{12}_\Lambda$B GS binding
energy at $11.37\pm0.06$ MeV(ref.\cite{hypmass}), the associate
statistical error and the composite nature ($2^-$/$1^-$) of the state
may cause the peak center to deviate from this value (ref.\cite{motoba}). So in the
calibration process, the peak position is a adjustable parameter and
the calibraion is carried out in a iterative way, the GS peak center
is adjusted after each iteration according to the actual fitted
position of calculated $^{12}_\Lambda$B GS distribution, until an
minimum Chisquare is reached.

The procedure is:

\begin{enumerate}
\item Calculate missing mass using the existing angle matrix and the initial
(un-calibrated) momentum matrix. From CH$_2$ target runs, select
$\Lambda$,$\Sigma^0$ events, from Carbon target runs, select
$^{12}_\Lambda$B GS events from a window located around the center
of the missing mass peaks.

Inside the 5mm thick CH$_2$ target, the energy loss and
radiative processes are significant. It shifts the missing
mass distributions and forms a tail.  The width of the event
window has to be selected appropriately, because a too wide
window will dilute the sample with too many radiated events while a too
narrow window will cause a insufficient statistics in the fitting.

\item Define  Chisquare as the sum of squared mass differences $\Delta
m_i^2$ between the
calculated mass and the PDB values or the initial assumed
$^{12}_\Lambda$B GS binding energy:
$$ \chi^2=\sum {w_i\Delta m_i^2}$$
where $w_i$ is the relative weight of $\Lambda$,$\Sigma$ and
$^{12}_\Lambda$B GS events. 

\item $\chi^2$ is a function of missing mass $emiss$,  thus a
  composite function of e' and
kaon momenta $\delta p_e$ and $\delta p_k$:
$$\chi^2=f(emiss(\delta p_e,\delta p_k)$$
Minimizing $\chi^2$ by by a Nonlinear Least 
Square method to optimize the Enge and HKS momentum
reconstruction matrices(ref.\cite{nls}).

\item fit the actual peak center of the calculated  $^{12}_\Lambda$B GS
missing mass distribution. Using this value as the new
$^{12}_\Lambda$B GS binding energy.  Go back to first step. Iteration
until an minimum Chisquare is reached.   

\end{enumerate} 

\section{Kinematics calibration}
The purpose of kinematics calibration is to find the offsets of beam
energy, Enge and HKS central momentum from nominal values. The method
is also to use the known masses of $\Lambda$,$\Sigma^0$ and
$^{12}_\Lambda$B ground state, same as momentum calibration.  The
$\chi^2$ is defined the same way as in momentum calibration. But for
kinematics calibration, the $\chi^2$ is a function of beam energy offset $\Delta
E_{e0}$, e' central momentum offset $\Delta p_{e'}$ and kaon momentum
offset $\Delta p_k$, corresponding to the zero order terms in the
momentum matrices. At correct offsets,
this $\chi^2$ will be minimized.

If we define sum of kinematics offsets as:

$$\Delta p_{kin}=\Delta E_{e0} - \Delta p_{e'}-\Delta p_k$$.

The $\chi^2$ will dominantly depend on $\Delta p_{kin}$. Correct
offsets can be found by scanning on $\Delta p_{kin}$ over all possible
values  to locate the minimum Chisquare.   

\section{Raster Correction}

To avoid burning the CH$_2$ target, beam will be rastered
5mm$\times$5mm on the target by a pair of bending magnets.  If the raster
effect is ignored, a direct reconstruction with matrices
obtained for point target gives much worse resolution, for example,
the $\delta p$ resolution for HKS is $1.16\times10^{-4}$(rms) without
raster, it becomes $6.64\times10^{-4}$ with 5mm$\times$5mm raster. 

To do raster correction, the beam positions on target need to be
determined by raster magnets X and Y current event by event. For
rastered beam, The reconstructed target quantities can be expressed as
functions of focal plane quantities and beam positions at target. It
can be written as, for example:

$$\delta p = f(xfp, xpfp, yfp, ypfp)+g(xtar, ytar)$$

\begin{figure}
\begin{center}
  \subfigure
{\includegraphics[width=7cm]{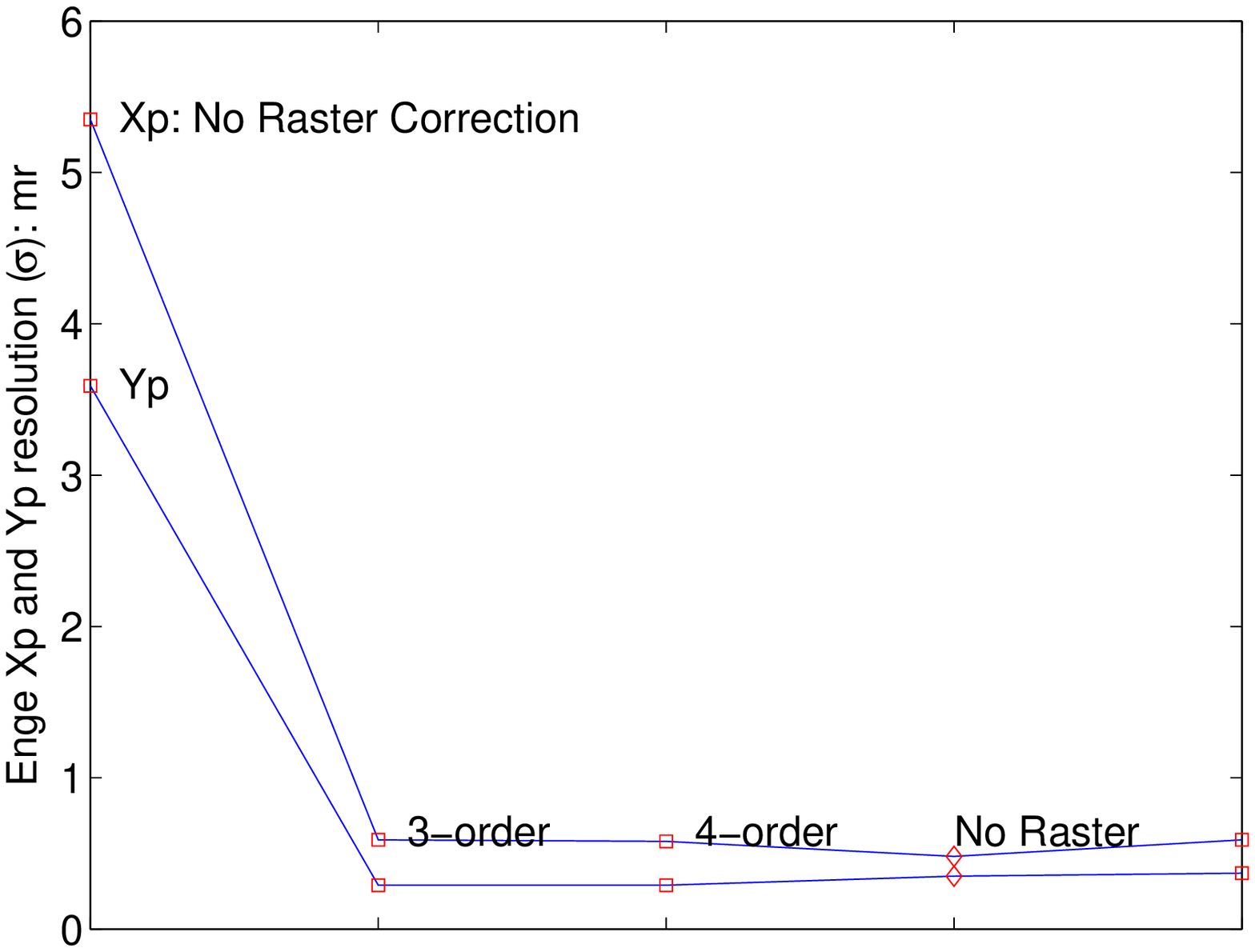}}
  \subfigure
{\includegraphics[width=7cm]{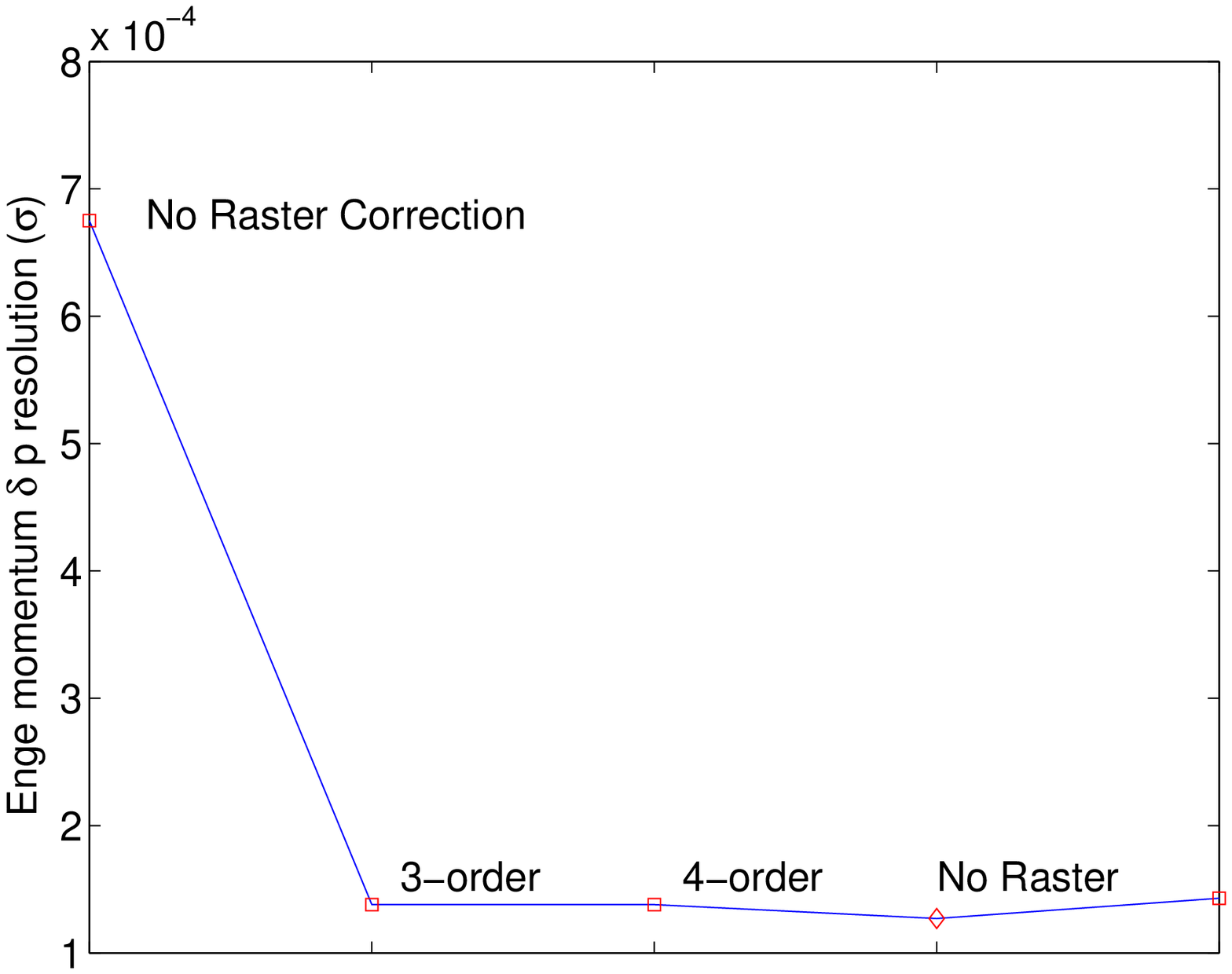}}
\end{center}
\caption{Comparison of Enge angle(left) and momentum(right) reconstruction resolution with
rastered beam,with and without raster correction. Both resolutions are
shown when the correction function $g$ fitted as a 3-order and 4-order
polynomials. The rightmost point is the resolution after raster
correction with wrong Enge optics.}
\label{fig:engeres_ras}
\end{figure}

\begin{figure}
\begin{center}
\includegraphics[width=7cm]{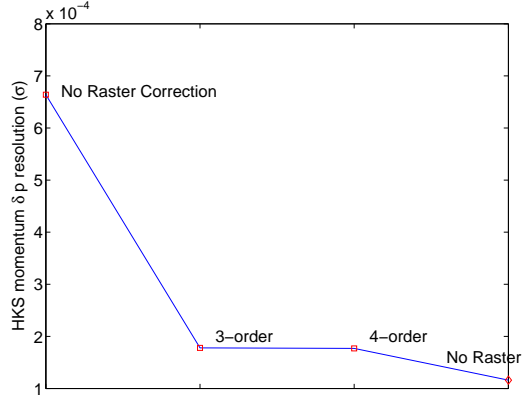}
\end{center}
\caption{Comparison of HKS momentum reconstruction resolution with
rastered beam,with and without raster correction.}
\label{fig:hksdp_ras}
\end{figure}

We seperated the dependence on beam positions from focal plane
quantities so that the raster correction function $g$ depends only on beam position.  It
can be fitted from initial spectrometer optics. Even if the actual
optics of the spectrometer system may deviate from the initial optics, its effect on raster correction function $g$ is negligible so we can
still use the initial function form (fig.\ref {fig:engeres_ras} and
fig.\ref {fig:hksdp_ras}).

\section {Iterative Calibration}

It is obvious that the individual calibration procedure mentioned
above actually are correlated. For example, the calibration of target
angles need the input of particle momentum while to extract missing
mass, we need to know the scattering angle. Thus the individual
calibraion procedures should be carried out in an iterative way until
convergence is reached. The general procedure will be:

\begin{enumerate}
\item Separate Enge and HKS sieve slit calibration to obtain angle
reconstruction matrices using original momentum reconstruction
matrix. 

\item Raster correction for CH$_2$ target data using existing raster
correction matrix.

\item Kinematics calibration to find deviation of beam energy, Enge
and HKS central momenta from their nominal values. The
central angles of the spectrometers are not used in the missing mass
calculation and they are not needed for fixed-angle spectrometer.

\item Two arm coupled momentum calibration to obtain momentum
reconstruction matrix. 

\item After getting the calibrated momentum matrix, go back to first step 
again to do SS calibration using the new momentum matrix. Iterate until best
resolution is reached.

\end{enumerate}

\section{Simulation of the Calibration Procedure}
Simulated events are used to test the calibration method. The
simulation take account of target physics processes, spectrometer
optics and detector resolution.

The simulation of target processes is adapted from SIMC (ref.\cite{simc}), a
physics simulation program for Jefferson Lab Hall C basic
equipment. The ionization energy loss, Bremsstrahlung, multiple
scattering and beam spread effects are included in the simulation. The
simulation uses 5mm thick CH$_2$ target and 0.1 g/cm$^2$ Carbon
target.

Simulated events from target are send through The
spectrometers. RAYTRACE (ref.\cite{raytrace}) is used to simulate events in the
electron arm. A measured magnet field map is used to track particles
through kaon arm by GEANT.  Only optical properties (no physical
processes) are considered in this step.

At focal plane of the spectrometers, the positions and angles are
smeared according to detector resolution and multiple
scattering. For Enge, the resolutions ($\sigma$) are: 86 $\mu$m (x),
0.7 mr(xp), 210 $\mu$m(y), 2.8 mr(yp). For HKS, they are: 160 $\mu$m
(x and y), 0.33 mr (xp and yp). 

The simulated focal plane events are then reconstructed back to the
target using backward reconstruction matrix. The beam energy and
particle momentum are corrected by the average energy loss in the
target. The corrections are 0.4923 MeV (beam energy), 0.6015 MeV (e'
momentum), 0.4890 MeV (kaon momentum) for CH$_2$ target, 0.0820 MeV
(beam energy),
0.1026 MeV (e' momentum), 0.0866 MeV (kaon Momentum) for Carbon
target. 

\begin{figure}
\begin{center}
\includegraphics [width=15cm]{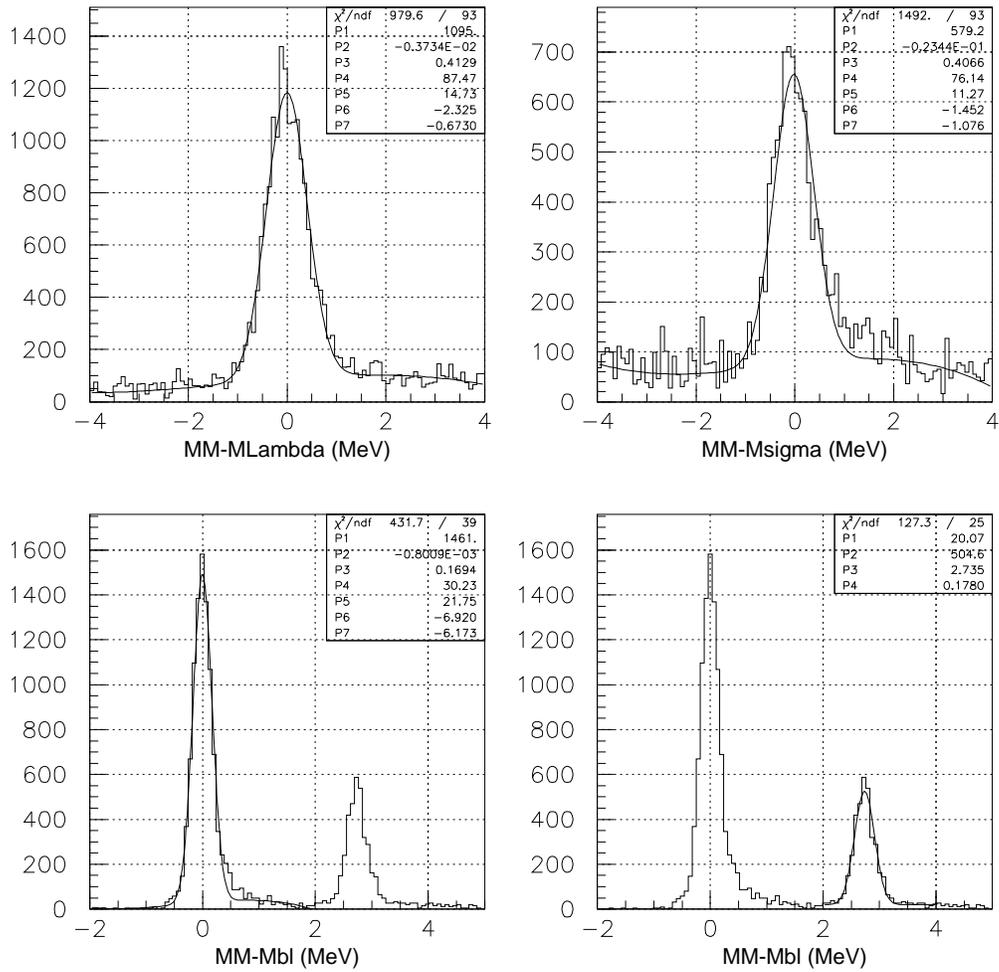}
\end{center}
\caption{Simulated $\Lambda$, $\Sigma^0$ and $^{12}_\Lambda$B missing mass
spectra with Carbon QF background from CH$_2$ target.}
\label{fig:mmco}
\end{figure}

The simulated $\Lambda$, $\Sigma^0$ and $^{12}_\Lambda$B missing mass
spectra are shown in fig.\ref{fig:mmco}. The predicted $1^-$ state of
$^{12}_\Lambda$B at 2.73 MeV is also included in the simulation for
comparison. The quasi-free carbon background on CH$_2$ target is
simulated with $\Lambda$ and $\Sigma^0$ events by a assumed S/B ratio
of 6:1 under $\Lambda$ peak. The $^{12}_\Lambda$B GS has a missing
mass resolution of 397 KeV (FWHM) with correct optics.

\begin{figure}
\begin{center}
\includegraphics[width=15cm] {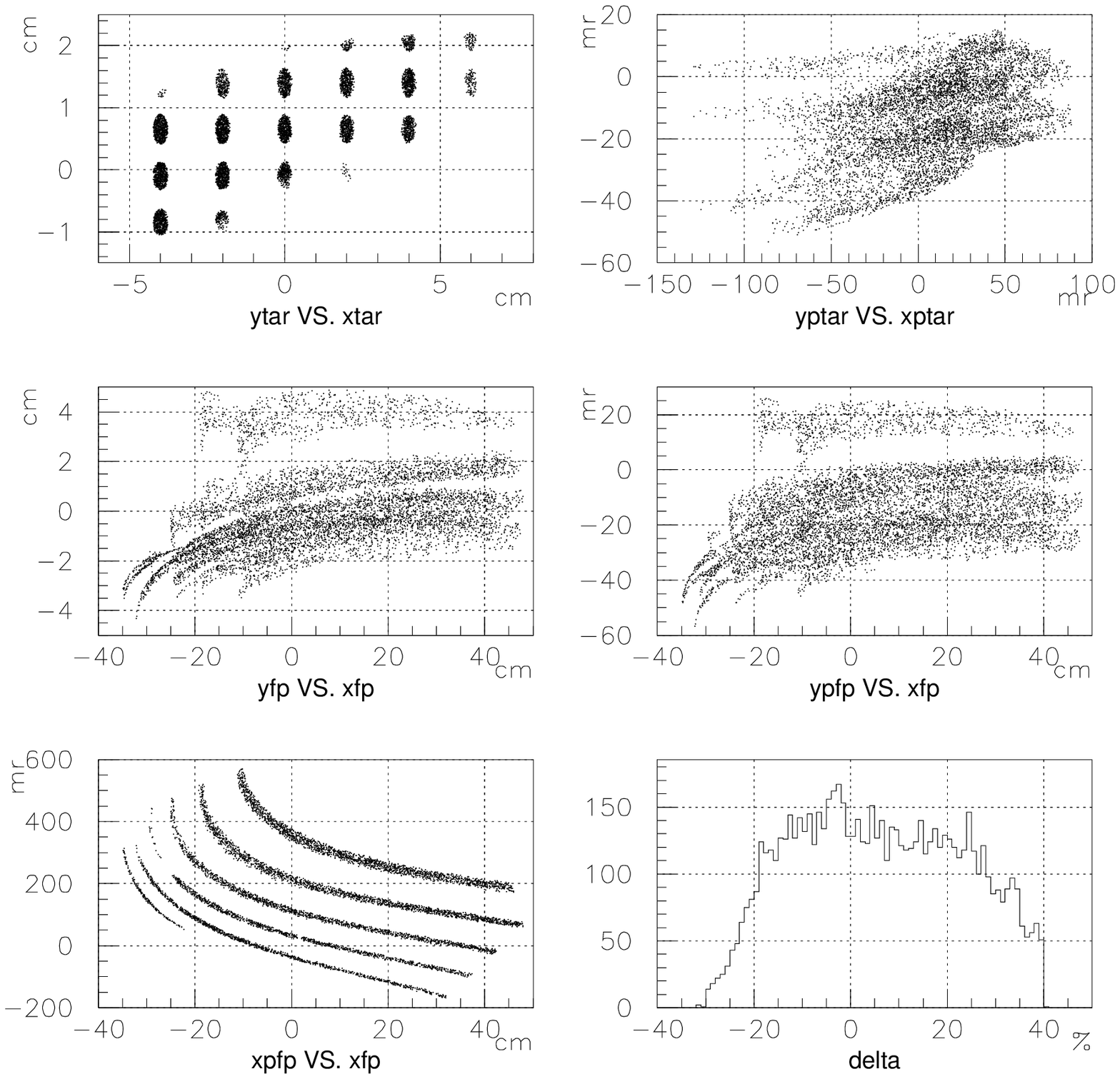}
\end{center}
\caption{Simulated Enge focal plane sieve slit correlation and hole patterns.}
\label{fig:engefpss1}
\end{figure}

\begin{figure}
\begin{center}
\includegraphics[width=15cm] {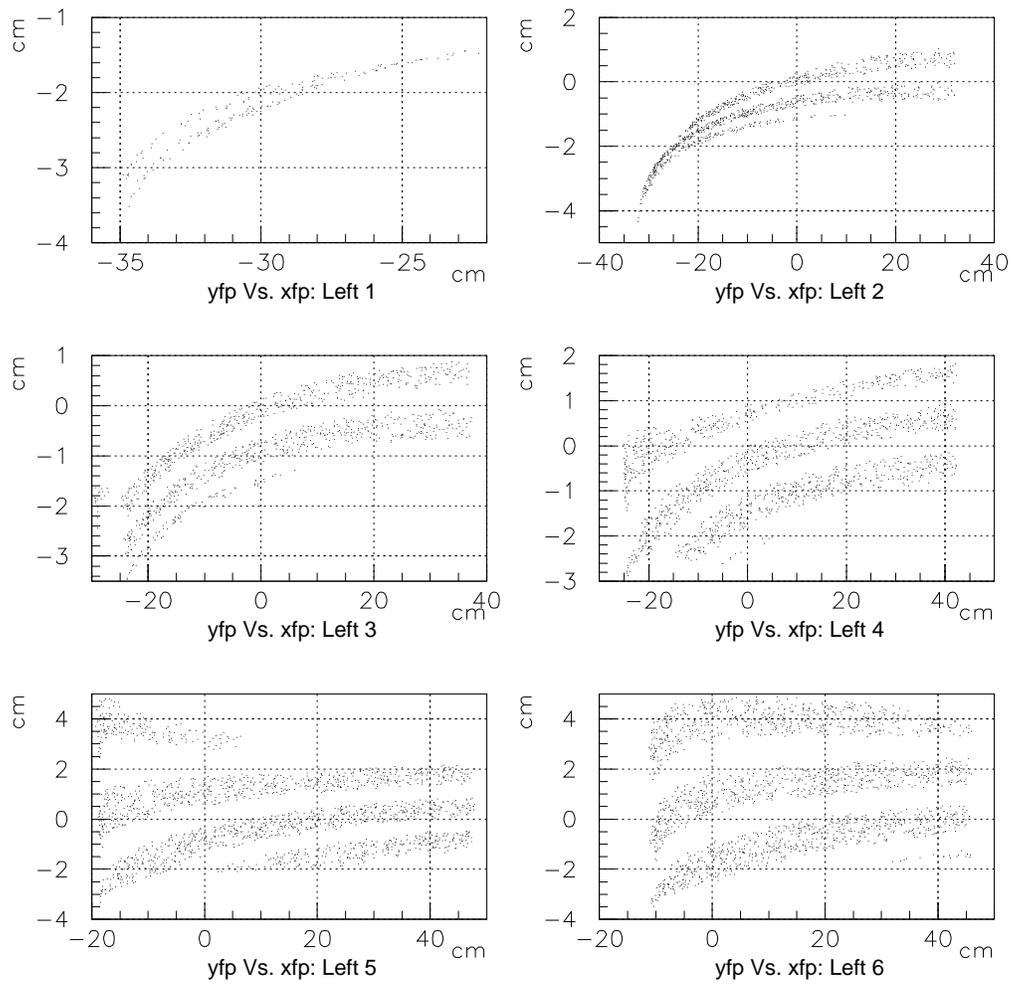}
\end{center}
\caption{Simulated Enge focal plane $yfp$ vs. $xfp$ correlations for each X-column of sieve slit holes.}
\label{fig:engefpss2}
\end{figure}

The simulated Enge focal plane sieve slit correlation patterns are
shown in fig.\ref{fig:engefpss1}. Fig.\ref{fig:engefpss2} shows the
yfp vs. xfp correlations for each X-column of sieve slit holes.

\begin{figure}
\begin{center}
\includegraphics[width=15cm] {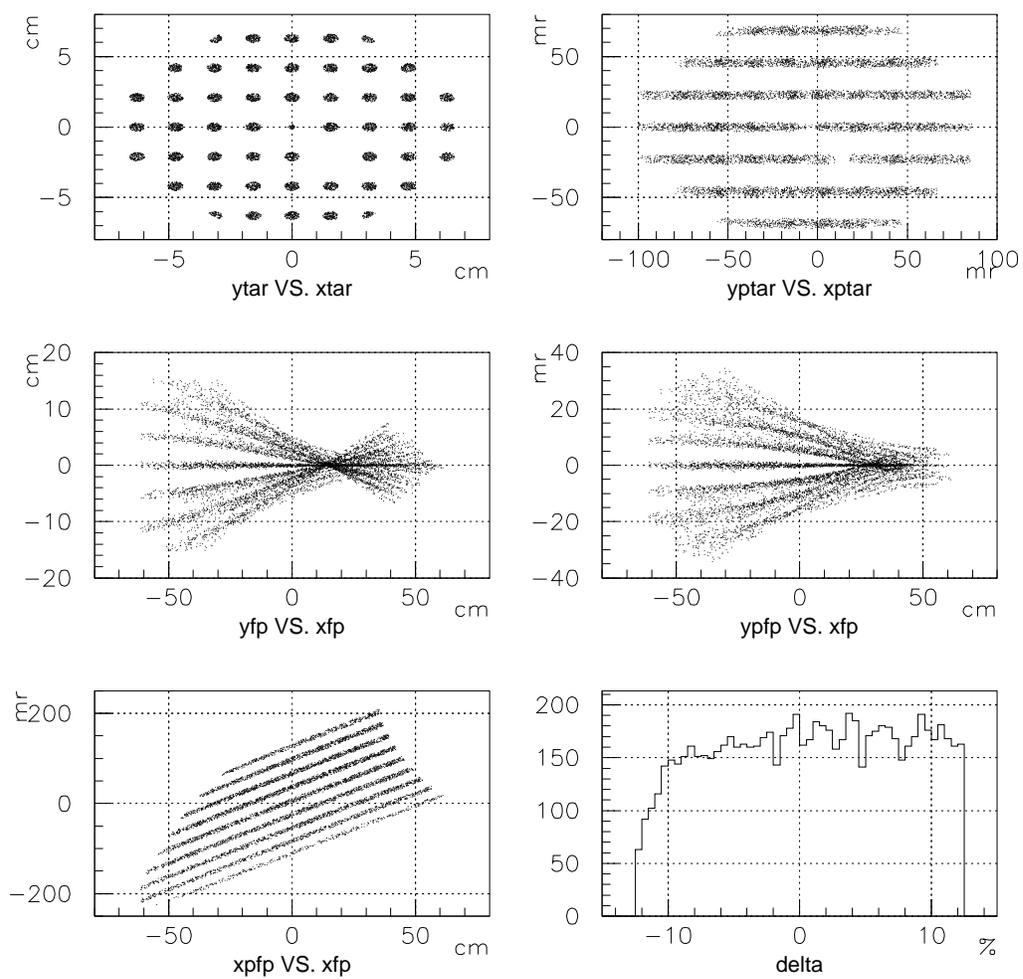}
\end{center}
\caption{Simulated HKS focal plane sieve slit correlation and hole patterns.}
\label{fig:hksfpss}
\end{figure}

The simulated HKS focal plane sieve slit patterns are
shown in fig.\ref{fig:hksfpss}.

\begin{figure}
\begin{center}
\includegraphics[width=15cm] {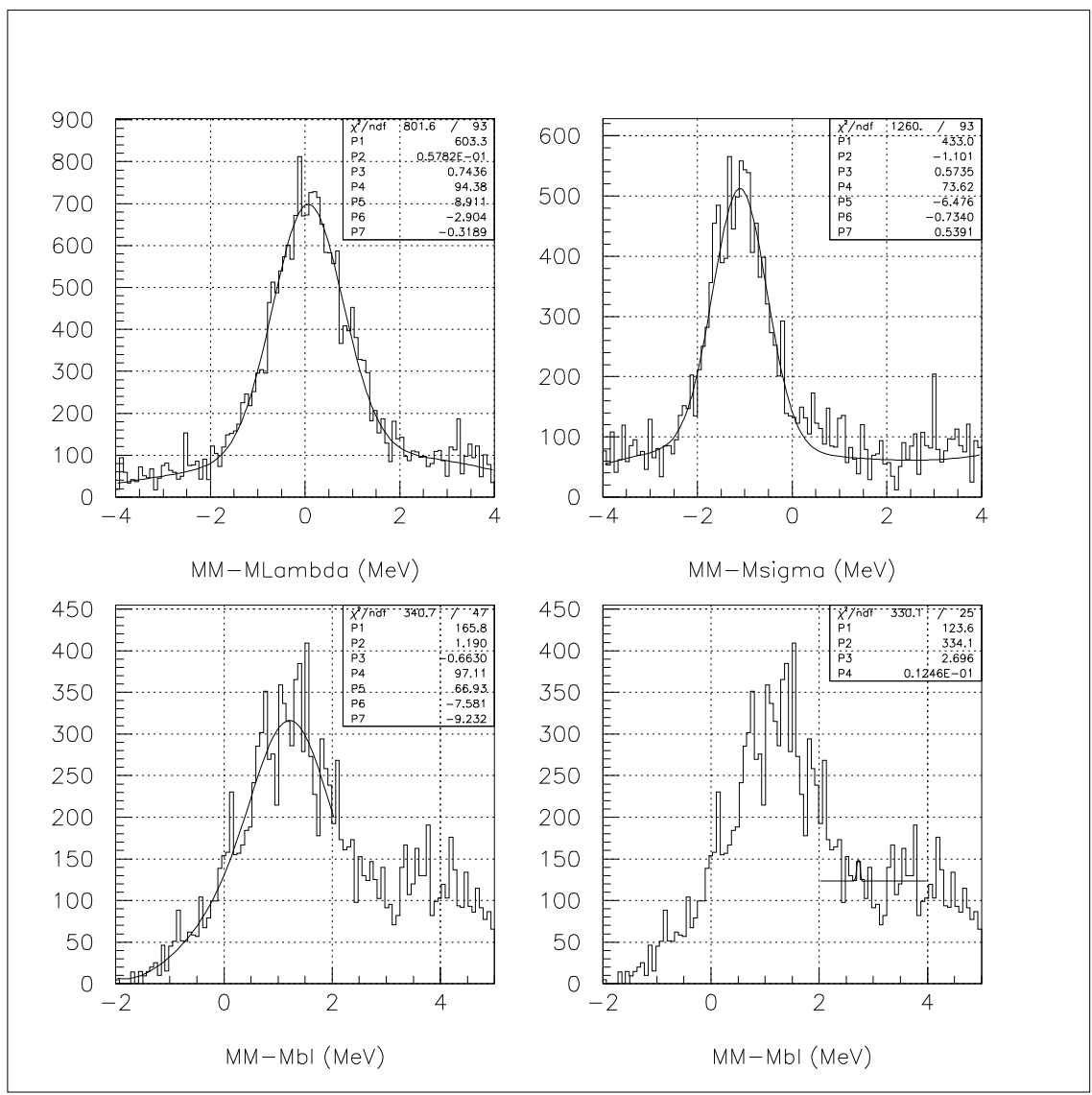}
\end{center}
\caption{Reconstructed $\Lambda$, $\Sigma^0$ and $^{12}_\Lambda$B missing mass
spectra with wrong optics before the calibration.}
\label{fig:mmwonc}
\end{figure}

\begin{figure}
\begin{center}
\includegraphics[width=15cm] {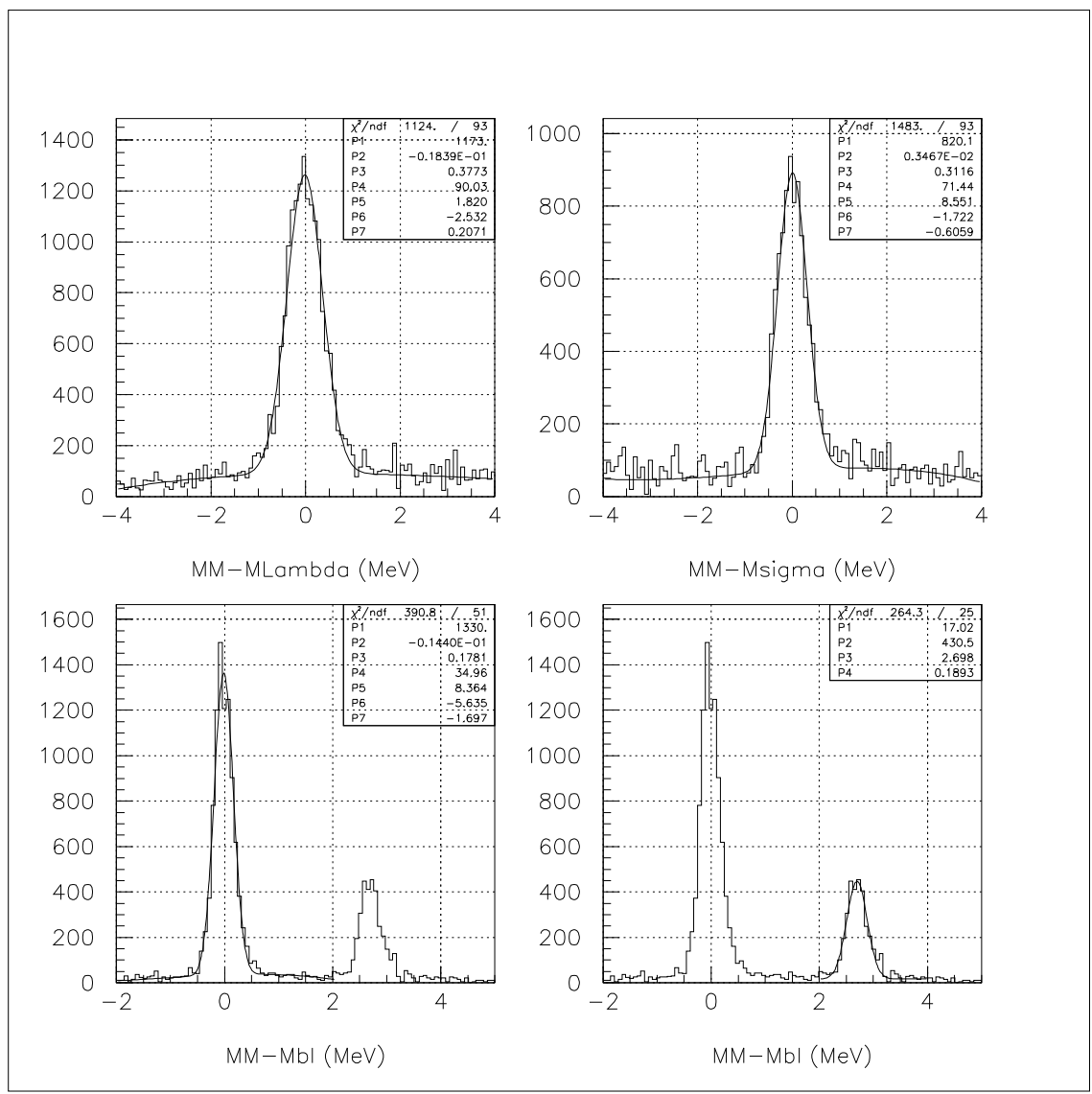}
\end{center}
\caption{Reconstructed $\Lambda$, $\Sigma^0$ and $^{12}_\Lambda$B missing mass
spectra with optical matrices calibrated.}
\label{fig:mmwoac}
\end{figure}

To test the calibration method by simulation, We intentionally use wrong
spectrometer optics in reconstruction. The splitter field is changed
from nominal value of 1.546 Tesla to 1.550 Tesla in forward simulation
for e' events, but still using the nominal 
reconstruction matrix in reconstruction. In addition, the beam energy,
electron and kaon arm central momenta have offsets from nominal
values, they are 1.2 MeV, -0.7 MeV and -0.9 MeV respectively. The
missing mass distributions before the calibration is shown
in fig. \ref{fig:mmwonc}. The missing mass distributions after the
calibration is in fig. \ref{fig:mmwoac}. The width of the
$^{12}_\Lambda$B GS improved from 1.56 MeV (FWHM) to 0.418 MeV (FWHM)
after the calibration, and the center of GS peak is within 20 keV of
the expected GS mass obtained by emulsion experiment.  

\section{Summary}

With the use of a dipole magnet on target, the normal spectrometer
calibration procedure has to be modified. We have developed a new
method for calibration of the reconstructions for target angle,
momentum and kinematics. Starting from a initially inaccurate optics,
by iteration of these calibration procedures, the spectrometer optics
can be calibrated to the required resolution.

\end{document}